\documentclass[preprint,aps,tightenlines,showpacs,nofootinbib]{revtex4}
\usepackage{latexsym}
\usepackage[dvips]{graphicx}
\@addtoreset{equation}{section}

\newcommand{\beq}{\begin{eqnarray}}
\newcommand{\eeq}{\end{eqnarray}}
\newcommand{\eq}{eqnarray}

\newcommand{\al}{{\alpha}}
\newcommand{\be}{{\beta}}
\newcommand{\ci}{\cite}
\newcommand{\ga}{{\gamma}}

\newcommand{\ep}{{\epsilon}}

\newcommand{\de}{{\delta}}

\newcommand{\La}{{\Lambda}}
\newcommand{\m}{{\mu}}
\newcommand{\n}{{\nu}}

\newcommand{\om}{{\omega}}
\newcommand{\Om}{{\Omega}}
\newcommand{\pa}{{\partial}}
\newcommand{\no}{{\nonumber}}
\newcommand{\f}{\frac}
\newcommand{\ra}{\rightarrow}
\newcommand{\lra}{\leftrightarrow}

\newcommand{\new}{\hbox{\scriptsize new}}
\newcommand{\ren}{\hbox{\scriptsize ren}}

\newcommand{\ther}{thermodynamics }

\newcommand{\temp}{temperature }

\newcommand{\hb}{\hat{\be}}
\newcommand{\hO}{\hat{\Om}}

\newcommand{\na}{\nabla}
\newcommand{\Am}{{\cal A}_-}
\newcommand{\Ap}{{\cal A}_+}
\newcommand{\dAm}{\de {\cal A}_-}
\newcommand{\dAp}{\de {\cal A}_+}
\begin{document}

\preprint{hep-th/0611048}

\title{Notes on the Area Theorem}

\author{Mu-In Park\footnote{E-mail address: muinpark@yahoo.com}}

\affiliation{ Center for Quantum Spacetime,  Sogang University,
Seoul 121-742, Korea \footnote{Present address: Research Institute
of Physics and Chemistry, Chonbuk National University, Chonju
561-756, Korea}}
\begin{abstract} Hawking's area theorem can be
understood from a quasi-stationary process in which a black hole
accretes $positive$ energy matter, {\it independent of the details
of the gravity action}. I use this process to study the dynamics of
the {\it inner} as well as the outer horizons for various black
holes which include the recently discovered exotic black holes and
three-dimensional black holes in higher derivative gravities as well
as the usual BTZ black hole and the Kerr black hole in four
dimensions. I find that the area for the inner horizon ``can
decrease'', rather than increase, with the quasi-stationary process.
However, I find that the area for the outer horizon
``never decrease'' such as the usual area theorem still works in our
examples, though this is quite non-trivial in general.
There exists an instability problem of the inner horizons
but
it seems that the instability is not important in
my analysis.
I also find a {\it generalized} area theorem by combining those of
the outer and inner horizons.
\end{abstract}

\maketitle

\newpage

\section
{Introduction}

Hawking's area theorem states that the total surface area of the
outer (event) horizon, ${\cal A}_+$, will never decrease in any
``classical'' processes involving black holes \ci{Hawk:71}. 
This is usually proved by considering the Raychaudhuri's equation
with several basic assumptions like as the stability of the
solution, as well as the cosmic censorship and an appropriate energy
condition with the field equations. But there exits a more physical
derivation of the theorem by considering a quasi-stationary process
in which the area law arises directly from the first law of
thermodynamics for any gravity theories, as far as the matters
satisfy the null energy condition \cite{Jaco:95}.

Recently the dynamics of the inner horizon became important in
several contexts
\ci{Krau:06,Solo:06,Saho:06,
Carl:91,
Bana:98,
Park:0602,Park:0608,Park:0609}. In some cases, the area of the inner
horizon, ${\cal A}_-$, appears in the black hole entropies
\ci{Solo:06,Saho:06,Carl:91,
Bana:98}
and in some other cases, the temperatures and the chemical
potentials for the inner horizon appear in the first law
\ci{Park:0602, Park:0608,Park:0609}. In these systems, it does not
seem to be straightforward to apply the Raychaudhuri's equation to
study the dynamics of the horizons, due to the instability of the
inner horizons which is suggested from the diverging energy-momentum
for the massless test fields at the inner (Cauchy) horizon
\ci{Simp:73}. Moreover, Poisson and Israel have shown how the
back-reaction effect of the diverging (energy-momentum) perturbation
on the Cauchy horizon produces a curvature singularity and an
unbounded inflation of the internal gravitational-mass parameter,
called ``mass inflation'' \ci{Pois:89}. 
However, in a more detailed analysis, Ori \ci{Ori:91},
has found that despite the curvature and mass-inflation
singularities, the associated metric perturbations are finite and
the metric tensor is nonsingular: Actually, the metric perturbations
are finite and small to ``all'' orders in the perturbation
expansion, even though their gradients (and the curvatures) diverge
at the Cauchy horizon. This suggests that the perturbation approach
may be applied to the inner horizon also. In this sense, it seems
that the instability is not important in these cases such as the
approach with the quasi-stationary process can be still used to get
some relevant information about the horizon dynamics.
Moreover, in the three-dimensional case which I am discussing in
this paper, the situation becomes better: The curvature scalars are
{\it insensitive} to the mass-inflation singularity, i.e., the
curvature scalars, as well as the metric, are finite \ci{Chan:96}.
And also, this seems to be consistent with the results in a quite
different approach \ci{Hayw:94}. 

In this paper I show, following the quasi-stationary approach, that
the area for the inner horizon ``can decrease'', as well as the
usual area-increasing theorem for the outer horizon, regardless of
some exotic behaviors in various situations
\ci{Krau:06,Solo:06,Saho:06,
Carl:91,
Bana:98}. 

The plan of this paper is as follows. In the next section, I
consider the BTZ-type black holes in
various gravity actions and demonstrate the Hawking's area
increasing theorem for the outer horizon and some possible area
decreasing for the inner horizon with the quasi-stationary
perturbation. I also consider a generalized area theorem by
combining those of the outer and inner horizons.
I finally consider the Kerr black hole in four dimensions and study
its area theorem. Though there are some differences in the details,
the similar behaviors of the horizon dynamics under the
quasi-stationary processes are obtained.
I conclude with some remarks.


\begin{section}
{The area theorem from quasi-stationary process}
\end{section}
In this section, I study the area theorem for the BTZ-type black
holes in various gravity theories and the Kerr black hole, from the
quasi-stationary process. Since the they are related by the same
geometric effect, I first study the usual BTZ black hole in Einstein
gravity for definiteness and then briefly discuss its application to
other various BTZ-type black holes. I finally study the Kerr black
hole in four dimensions.\\

\begin{subsection}
{The BTZ black hole
}
\end{subsection}

The (2+1)-dimensional gravity with a cosmological constant
$\La=-1/{l}^2$ is described by the action on a manifold ${\cal M}$ [
omitting some boundary terms ]
\begin{\eq}
\label{EH} I_{g}=\frac{1}{16 \pi G} \int_{\cal M} d^3 x \sqrt{-g}
\left( R +\frac{2} {{l}^{2}}~\right)+I_{matter},
\end{\eq}
where $I_{matter}$ is a matter action whose details are not
important in this paper.

The equations of motion for the metric are given by
\begin{\eq} \label{eom}
R^{\mu \nu}-\f{1}{2} g^{\m \n} R -\f{1}{{l}^2} g^{\m \n} =8 \pi G
~T^{\m \n}_{matter}
\end{\eq}
with the matter's energy-momentum tensor
 $T^{\m \n}_{matter}$.
These equations imply that the solution have a constant curvature
$R=-6/l^2$ outside the matter. There are many non-trivial solutions
with matters (see Refs. \ci{Carl:95,Park:04} and references
therein). In this paper, it is enough to consider a vacuum black
hole solution, known as Banados-Teitelboim-Zanelli (BTZ) solution,
which is given by the metric \cite{Bana:92}
\begin{eqnarray}
\label{BTZ}
 ds^2=-N^2 dt^2 +N^{-2} dr^2 +r^2 (d \phi +N^{\phi}
dt)^2
\end{eqnarray}
with
\begin{eqnarray}
\label{BTZ:N}
 N^2=\frac{(r^2-r_+^2) (r^2-r_-^2)}{{l}^2 r^2},~~
N^{\phi}=-\frac{r_+ r_-}{{l} r^2}.
\end{eqnarray}
Here, $r_+$ and $r_-$ denote the outer and inner horizons,
respectively. The mass and angular momentum of the black hole are
given by
\begin{eqnarray}
\label{mj:bare}
 m=\frac{r_+^2 +r_-^2}{8 G {l}^2},~~j=\frac{2 r_+ r_-}{8G l },
\end{eqnarray}
respectively. The radii of inner and outer horizons are given by
\begin{\eq}
\label{horizon}
 r_{\pm} =l \sqrt{ 4G m \left[1 \pm \sqrt{1-(j/ml)^2}\right]}.
\end{\eq}
Note that the mass and angular momentum parameters, which are
$positive$ semi-definite, $m \geq 0, j \geq 0$, satisfy the usual
mass and angular momentum inequality
\begin{\eq}
\label{ieq:BTZ}
 m\geq j/{l}
\end{\eq}
in order that the horizons of (\ref{horizon}) exist or the conical
singularity is not naked, with the equality for the extremal black
hole having the overlapping inner and outer horizons.

Then, it is easy to see that the quantities in (\ref{mj:bare})
satisfy the following identity, for an ``arbitrary'' variation $\de$
and the area of the outer (event) horizon, ${\cal A}_+=2 \pi r_+$,
\begin{\eq}
\label{first:BTZ}
 \de m-\Om_+ \de j =\f{T_+}{4 \hbar G} ~\de {\cal
A}_{+},
\end{\eq}
where the coefficients $\Om_+$  and $T_+$ are the angular velocity
and Hawking temperature for the outer horizon $r_+$
\begin{eqnarray}
\label{T:BTZ}
 \Omega_+=\left.-N^{\phi} \right|_{r_+}=\frac{r_-}{l
r_+},~~T_+=\left. \frac{\hbar \kappa}{2 \pi}
\right|_{r_+}=\frac{\hbar (r_+^2 -r_-^2)}{2 \pi l^2 r_+},
\end{eqnarray}
respectively, with the surface gravity function $ \kappa={2}^{-1}
{\partial N^2}/{\partial r}$.

Given the identification of the temperature $T_+$, Eq.
(\ref{first:BTZ}) has a natural interpretation as the first law of
\ther with the entropy
$S_{BH}={{\cal A}_+}/{4 \hbar G},$
by choosing the undetermined integration constant to be zero
\ci{Beke:73}. Note that there would be no justification of the
entropy as $S_{BH}$ unless one can prove the second law of
thermodynamics, i.e., the area-increasing theorem, in any dynamical
process; without this justification, $T_+$ as the temperature can
not be justified, either. Usually this is proved by considering the
Raychaudhuri's equation with the null energy condition for the
matter's energy-momentum tensor \ci{Hawk:71}.
Here,
I consider a quasi-stationary process approach to get the theorem,
for the generality of the derivation \ci{Jaco:95}. By a
``quasi-stationary'' process I mean one in which the background
spacetime is only slightly perturbed. But, I will not restrict to
the perturbations by the infalling matters only but it could be some
classical perturbations in the metric itself. For this
quasi-stationary process one can use the first law (\ref{first:BTZ})
for ``small'' parameters $\de$, though it is an ``exact'' formula
for an arbitrary $\de$, within the given solution.

To this end, I note that the inequality (\ref{ieq:BTZ}) provides a
restriction on the quasi-stationary process such as (\ref{ieq:BTZ})
is not violated by the perturbed mass and angular
momentum\footnote{Actually this can be also understood dynamically,
by studying the capture of the test particles by the black holes
\ci{Wald:74}.}:
\begin{\eq}
\label{drest:BTZ}
 \de m \geq \de j/l \geq 0.
\end{\eq}
This restriction  corresponds to the ``dominant'' energy condition
if I consider the perturbation from the infalling of $positive$
energy matters, rather than the ``null'' energy condition in the
usual derivation \ci{Hawk:71,
Jaco:95}; the difference is immaterial in the final results but the
analysis can be more manifest with this restricted condition. Then I
find, for $T_+>0$ and $\Om_+>0$ (or equivalently $r_+>r_->0$), that
\begin{\eq}
\label{area:BTZ}
\de {\cal A}_+&=&\f{4 \hbar G}{T_+} ~(\de m-\Om_+ \de j) \no \\
 &\geq& \f{4 \hbar G}{T_+} ~ \left( 1-\f{r_-}{r_+}
 \right) \de m \geq 0,
\end{\eq}
i.e., the non-decreasing area ${\cal A}_+$ for the outer horizon.
This demonstrates the usual Hawking's area theorem
\ci{Hawk:71}
for the ``restricted'' class of perturbation
with (\ref{drest:BTZ}). But note that $\de {\cal A}_+$ can not be
determined for the extremal black hole background with $T_+=0$. I
also note that (\ref{first:BTZ}) or (\ref{area:BTZ}) shows an
interesting complementary between the variations in quantities of
black hole as measured at the asymptotic infinity, $\de m$ and $\de
j$, with the variation of a geometric property of the horizon $\de
{\cal A}_+$.

In the usual context of (\ref{first:BTZ}) and (\ref{area:BTZ}),
there is no chance to get the information about the inner horizon
and also the positive temperature $T_+$ is crucial for the validity.
But there are several other variations of (\ref{first:BTZ}) and
(\ref{area:BTZ}), depending on the systems, which generalize the
usual context and can give the information about the inner horizon.
In this BTZ black hole case first, the key observation for this new
analysis is to consider another identity which differs from
(\ref{first:BTZ}) \ci{Park:0608}, for an arbitrary variation $\de$,
\begin{\eq}
\label{id2}
 \de m-\Om_- \de j =\f{(r_-^2 -r_+^2)}{8 \pi l^2 G r_-} ~\de {\cal A}_-,
\end{\eq}
where the coefficient $\Om_-$ is the angular velocity of the inner
horizon $r_-$,
\begin{\eq}
\label{Om-} \Omega_-=\left.-N^{\phi} \right|_{r_-}=\frac{r_+}{l r_-}
\end{\eq}
and ${\cal A}_-=2 \pi r_-$ is the area of the inner horizon. This
new identity would be basically due to the symmetry between $r_+$
and $r_-$ in the metric (\ref{BTZ}) and actually this new identity
is unique. One $might$ consider (\ref{id2}) as the first law of
\ther for the inner horizon, by identifying
\begin{\eq}
\label{T-:BTZ}
 T_-=\f{\hbar{(r_-^2 -r_+^2)}}{{2 \pi l^2  r_-}}
\end{\eq}
and $S_-={{\cal A}_-}/{4 \hbar G}$ as the characteristic \temp and
entropy, respectively, but here there is ``no'' physical motivation
due to the fact that the entropy $S_-$ can not be justified as one
can see below; $T_-$ , which is $negative$ for $r_+>r_-$, should be
considered just as a coefficient in the formula (\ref{id2}) to
analyze the dynamics of ${\cal A}_-$.

Now, in order to study the dynamics of the inner horizon, let me
consider the quasi-stationary process again with the condition
(\ref{drest:BTZ}), as the above analysis for the outer horizon. But,
from the $negative$ coefficient, which being $T_-/4 \hbar G$, of the
right hand side in (\ref{id2}), I find\footnote{If I use the null
energy condition for the perturbation from the matter infalling
\ci{Jaco:95}, which is equivalent to $\de m -\Om_+ \de j \geq 0$, I
can obtain $\de {\cal A}_+ \geq 0$ ``generically'' but
(\ref{area:BTZ:inner}) changes as $\de {\cal A}_- \leq 8 \pi l^2 G
\de m/r_-$. So, the result is not much different from
(\ref{area:BTZ:inner}) but here $\de m$ needs not to be positive.}
that
\begin{\eq}
\label{area:BTZ:inner1}
\de {\cal A}_-&=&\f{4 \hbar G}{T_-} ~(\de m-\Om_- \de j) \\
\label{area:BTZ:inner}
 &\leq& \f{4 \hbar G}{T_-} ~ \left( 1-\f{r_+}{r_-}
 \right) \de m ,
\end{\eq}
for $r_+>r_-(>0)$. The ``upper'' bound of the area change $\de {\cal
A}_-$ is in contrast to the usual $lower$ bound as in
(\ref{area:BTZ}); actually the values of the upper bound and lower
bound are the same. This signals the unusual behavior of the inner
horizon and remarkably one can discover the ``non-increasing'' area,
rather than non-decreasing, for certain cases. In order to
demonstrate this, let me consider the particular process with
\begin{\eq}
\label{dm:particular:BTZ}
 \de m \geq 0,~~ \de j=0
\end{\eq}
which does not violate (\ref{ieq:BTZ}) and
(\ref{drest:BTZ})\footnote{This does not violate the null energy
condition either.}. Then, from (\ref{area:BTZ:inner1}), one finds
that, for $r_+ >r_-$, i.e., $T_- <0$,
\begin{\eq}
\label{area:BTZ:inner2}
\de {\cal A}_-&=&\f{4 \hbar G}{T_-}~\de m
\leq 0,
\end{\eq}
i.e., the ``non-increasing'' area $\Am$ for the inner horizon. Of
course, this does not mean that $\Am$ never increase since one can
find also some other appropriate $\de m$ and $\de j$, even within
(\ref{drest:BTZ}), such as $\dAm$ becomes positive: This can be
easily seen by noting that the upper bound of $\dAm$ in
(\ref{area:BTZ:inner}) can still be positive for the generalized
process with (\ref{drest:BTZ}), instead of the particular one
(\ref{dm:particular:BTZ}). But the result (\ref{area:BTZ:inner2})
only demonstrates that the inner horizon area, with the usual
appropriate factor $(4 \hbar G)^{-1}$, can ``not'' be considered as
an entropy since the second law of thermodynamics is not guaranteed
in general. This might depend on the black hole systems but this
seems to be generic, as far as the black holes in this paper are
concerned. For the extremal black hole background, $\dAm$ can not be
determined either as in $\dAp$.

As the final remark of this section, the two horizons merge by the
decreasing $\Ap$ and increasing $\Am$, contrary to (\ref{area:BTZ})
and (\ref{area:BTZ:inner2}), by considering $\de m=0, \de j >0$ for
example but this would not occur actually since the energy
conditions, like as the null energy condition as well as the
dominant energy condition, would be easily violated. Moreover, by
the definition of the quasi-stationary process, the causal structure
should not be changed such as $r_+$ would not cross $r_-$. Actually,
by comparing (\ref{area:BTZ}) and (\ref{area:BTZ:inner}), one finds
that $\dAm \leq \dAp$ ( equality for $r_+=r_-$ ) within the
condition (\ref{drest:BTZ}).\\

\begin{subsection}
{The exotic BTZ black holes}
\end{subsection}

Recently a number of unusual
black hole solutions, which are called the ``exotic BTZ black
holes'' have been found. These are (a) asymptotically anti-de Sitter
black holes in (2+1)-dimensional gravity for the case of a vanishing
cosmological constant with minimally coupled topological matter,
which is called `` BCEA '' gravity \cite{Carl:91}\footnote{Another
similar model has been recently studied, similar to the BCEA theory
but with zero-form and two-form matter fields. But, unlike the BCEA
case, the Noether charges of this solution vanish identically
\ci{Mann:05}. }, (b) constant curvature black holes in
(4+1)-dimensional anti-de Sitter space \cite{Bana:98},
and (c) BTZ-like black hole in gravitational Chern-Simons theory
\cite{Solo:06}. They have the following ``universal'' behaviors:

a. Their local metrices are the same as the BTZ black hole solution
(\ref{BTZ}), (\ref{BTZ:N}) \ci{Carl:91,
Solo:06}, or modulus a 2-sphere \ci{Bana:98}. 

b. Their masses and angular momenta are completely interchanged from
the ``bare'' quantities $m,j$ as
\begin{eqnarray}
\label{M_J}
 M=x j/l, ~~ J=x l m
\end{eqnarray}
with an appropriate coefficient $x$; $x=1$ in Ref. \cite{Carl:91},
$x$ is a fixed value of $U(1)$ field strength in Ref.
\cite{Bana:98}, 
and $x$ is proportional to the
coefficient of a gravitational Chern-Simons term in Ref.
\cite{Solo:06}. One remarkable result of (\ref{M_J}) is an anomalous
inequality in the mass and angular momentum
 $
 (l M)^2 -J^2 =x^2 [j^2 -(lm)^2] \leq 0
$
for any non-vanishing $x$, which shows an upper bound for the
mass-squared $M^2$ and a saturation for the extremal case in the
bare quantities $m^2=j^2/l^2$; this is the condition of the
existence of the horizons for the exotic black holes, as
(\ref{ieq:BTZ}) is for the BTZ black hole.

On the other hand, it has been known that these black holes have the
black hole entropy, which is related to the Wald's Noether entropy
\ci{Wald:93,Iyer:94}, as
\begin{eqnarray}
\label{BH_old} S_{exo}=x \frac{\Am}{4 G \hbar },
\end{eqnarray}
which depends on the $inner$ horizon area ${\cal A}_-$, rather than
the outer horizon's ${\cal A}_+$
\cite{Carl:91,
Bana:98,
Solo:06}. I note that this entropy satisfies the first law, which
descends from (\ref{id2}), with the usual Hawking temperature and
angular velocity as $T_+$ and $\Om_+$, respectively,
$
\de M- \Om_+ \de J=T_+ \de S_{exo}.
$ ; there is no other choice for the entropy, in the {\it usual}
context \ci{Solo:06,Carl:91,
Bana:98}.

But recently, I have pointed out that the second law of \ther is
questionable with the entropy (\ref{BH_old}) and the new entropy
formula
\begin{eqnarray}
S_{exo(\new)}=|x| \frac{\Ap}{4 G \hbar} \label{BH_new}
\end{eqnarray}
is needed such as the second law be guaranteed \ci{Park:0602}. This
new proposal is consistent with
an independent computation of the (statistical) entropy from a
conformal field theory (CFT) formula, known as the Cardy formula
\ci{Park:0602,Park:0608}. Now, the violation of the second law for
the entropy (\ref{BH_old}), with $x>0$, is evident from the ``same''
geometric effect as (\ref{area:BTZ:inner2}) with the condition
(\ref{drest:BTZ}), which does not depend on the details of the
actions and the definition of the mass and angular momentum; for
$x<0$, the second law might not be violated but the entropy becomes
negative. On the other hand, with the same argument, the second law
of the new entropy formula (\ref{BH_new}) is evident from
(\ref{area:BTZ}).\\

\begin{subsection}
{
With higher curvatures}
\end{subsection}

The (2+1)-dimensional gravity with the higher curvature terms and a
$bare$ cosmological constant $\La=-1/{l}^2$ can be $generally$
described by the action
\begin{\eq}
\label{Higher} I_{g}=\frac{1}{16 \pi G} \int_{\cal M} d^3 x
\sqrt{-g} \left( ~f(g^{\m \n}, R_{\m \n}, \nabla_{\m}) +\frac{2}
{{l}^{2}}~\right)+I_{matter},
\end{\eq}
where $f(g^{\m \n}, R_{\m \n}, \nabla_{\m})$ is an $arbitrary$
scalar function constructed from the metric $g^{\m \n}$, Ricci
curvature tensor $R_{ \m \n}$, and the covariant derivatives
$\nabla_{\m}$ \cite{Jaco:94,Iyer:94,Said:00}. This action is the
most generic, diffeomorphically invariant form in three dimensions
since there is no independent component of the Riemann tensor, due
to the vanishing Weyl tensor. The equations of motion, by varying
(\ref{Higher}) with respect to the metric, are
\begin{\eq}
\label{eom:Higher}
 \f{\pa f}{\pa g_{\mu \nu}}-\f{1}{2} g^{\m \n} f -\f{1}{{l}^2} g^{\m \n}
=8 \pi G \left(t^{\m \n}+T^{\m \n}_{matter} \right)
\end{\eq}
with the pseudo tensor
$
 t^{\m \n}=( \na^\n \na^\al {P_\al}^\m + \na^\m
\na^\al {P_\al}^\n -\Box P^{\m \n}-g^{\m \n} \na^\al \na^\be P_{\al
\be} )/{16 \pi G}  $
and $P_{\al \be} \equiv g_{\al \m} g_{\be \n} (\pa f/\pa R_{\m
\n})$. 

In the absence of the higher curvature terms, the BTZ solution,
(\ref{BTZ}), (\ref{BTZ:N}) is the ``unique'' black hole solution in
vacuum, due to the three-dimensional Birkoff's theorem \ci{Beat:04}.
However, in the presence of the higher curvature terms, it would not
be unique anymore since the higher curvature terms act like as the
matters with the energy-momentum $t_{\m \n}$ such as it would not be
the vacuum effectively: In the presence of the matters, we already
have several black hole solutions which modify the BTZ solution
non-trivially \ci{Chan:97,
Park:04}.

But, even in the presence of the generic higher curvature terms, the
BTZ solution can be still a vacuum solution since the $local$
structure would be ``unchanged'' by the higher curvatures. The only
effects would be some ``re-normalization'' of the bare parameters
$l,r_{\pm}$, and the Newton's constant $G$
\cite{Saho:06,Park:0609,Said:00}; and in this case, one finds $t^{\m
\n}=0$ trivially from $P_{\al \be}\propto g_{\al \be}$ for any {\it
constant-curvature} solutions. The renormalized cosmological
constant will be denoted by
$\La_{\ren}=-1/l^2_{\ren}$ and the function $l_{\ren}=l_{\ren}(l)$
depends on the details of the function $f$; but I shall use the same
notations $r_\pm$ in the renormalized frame also, for brevity. The
renormalized Newton's constant is given by
$G_{\ren}=\hO^{-1} G,$
with the conformal factor $\hO$, defined by
\begin{\eq}
\label{Om}
 \hO \equiv \f{1}{3} g_{\m \n} \f{\pa f}{\pa R_{\m \n}},
\end{\eq}
which being constant for any constant-curvature solutions
\cite{Said:00}. But note that this normalization factor is different
from that of the cosmological constant \footnote{For $f=R+a R^2+b
R_{\m \n}R^{\m \n}$ with some appropriate coefficients $a,~b$
\cite{Said:00}, the function $l_{\ren}=l_{\ren}(l)$ is given by $-6
l_{\ren}^{-2}=\left(-1 \pm \sqrt{1-24 (b-a)l_0^{-2}}\right)/(2
(b-a))$. But the conformal factor is given by
$\hO=1-(12a+4b)l^{-2}$.}. Now, due to the renormalization of the
Newton's constant, the original mass and angular momentum are
modified as
\begin{\eq}
\label{MJ:higher}
 M=\hO m,~~J=\hO j ,
\end{\eq}
respectively. Here, $m$ and $j$  represent the usual mass and
angular momentum for the metric (\ref{BTZ}) in the ``renormalized''
frame $m=(r_+^2 +r_-^2)/8 G l^2_{\ren},j=(2 r_+ r_-)/8G l_{\ren}$,
with the renormalized parameters $l_{\ren},r_\pm$, but with the bare
Newton's constant $G$, such as $m\geq j/l_{\ren} $ is valid still.
And it is important to note that $\hO$ is $not$ positive
definite\footnote{This means that the renormalized Newton's constant
$G_{\ren}$ can be negative (the bare Newton's constant $G$ is
assumed to be positive). But this does $not$ mean the
``antigravity'' since there are no gravitons which can mediate the
interactions between massive particles in three dimensional
gravities which are described by the action (\ref{Higher}): Here,
there is no a priori reason to fix the sign of the Newton's constant
and so both signs are allowed \cite{Dese:84}. However, this $can$
imply the antigravity, even without some $twistings$ due to a
``spinning'' source \ci{Dese:90}, when there are propagating
gravitons in the presence of a gravitational Chern-Simons term
\ci{Dese:82}.} such as the usual inequality for the mass and angular
momentum would not be valid in general,
but depends on the sign of $\hO$.

Then, from (\ref{MJ:higher}), (\ref{first:BTZ}), and the
identification of the temperature $T_+$ as (\ref{T:BTZ}), one has a
natural interpretation (\ref{first:BTZ}) as the first law, with the
entropy
\begin{eqnarray}
\label{S:Wald}
 S_W=\hO \frac{{\cal A}_+}{4 G \hbar},
\end{eqnarray}
which agrees with the Wald's entropy formula \cite{Said:00}.
From the same geometric, area-increasing effect, it is clear that
(\ref{S:Wald}) satisfies the second law for $\hO>0$.
However, for $\hO<0$, (\ref{S:Wald}) would not satisfy the second
law since it would ``decrease'' indefinitely, with the negative
values, as the outer horizon $r_+$ be increased.
Motivated by the statistical entropy computation, I have recently
proposed the new entropy formula \ci{Park:0609} which is an
increasing function of the area ${\cal A}_+$:
\begin{eqnarray}
\label{S:Wald(new)}
 S_{W(\new)}=|\hO| \frac{{\cal A}_+}{4 G \hbar}.
\end{eqnarray}
This satisfies the second law manifestly.

 On the other hand,
it is also easy to see that the area non-increasing behavior of
${\cal A}_-$ can be also demonstrated, as in the previous cases.\\

\begin{subsection}
{
With a gravitational Chern-Simons term}
\end{subsection}

In (2+1) dimensions, the following gravitational Chern-Simons (GCS)
term \ci{Dese:82} can be considered as a higher derivative
correction, other than the diffeomorphically invariant higher
``curvature'' corrections in (\ref{Higher}), to the Einstein-Hilbert
action in (\ref{EH}):
\begin{eqnarray}
\label{GCS}
 I_{GCS}=\frac{\hat{\beta} l}{64 \pi G} \int_{\cal M} d^3 x
~\epsilon^{\mu \nu \alpha} \left( R_{ab \mu \nu}
{\omega^{ab}}_\alpha +\frac{2}{3}~ {\omega^b}_{c_\mu} {\omega^c}_{a
\nu} {\omega^a}_{b\alpha} \right).
\end{eqnarray} Here, the spin-connection $1-form$
${\omega^a}_b={\omega^a}_{b \mu}
dx^{\mu},~\omega_{ab\mu}=-\omega_{ba\mu}$ is determined by the
torsion-free condition $d e^a +{\omega^a}_b \wedge e^b=0$ with the
dreibeins 1-form $e^a={e^a}_{\mu} dx^{\mu}$, and the curvature is
given by $R_{ab \mu \nu}=\pa_{\mu}\omega_{ab \nu}+{{\om_a}^c}_{\mu}
\om_{cb\nu}-(\m \lra \n)$. [ I take the same definitions as in Ref.
\cite{Krau:06} for the curvature 2-form $R_{ab}=(1/2)R_{ab\m
\n}~dx^{\m} \wedge dx^{\n}$ and the spin-connection 1-form
$\om_{ab}$. ]

The resulting equations of motion, by varying $I_{g}+I_{GCS}$, with
$I_g$ of (\ref{EH}), with respect to the metric are
\begin{\eq}
\label{eom:EHGCS}
 R^{\mu \nu}-\f{1}{2} g^{\m \n} R -\f{1}{l^2} g^{\m \n}
=\hat{\be} l C^{\m \n}+8 \pi G T^{\m \n}_{matter},
\end{\eq}
where the Cotton tensor $C^{\m \n}$ is defined by
$ C^{\mu \nu}= \epsilon^{\mu \rho \sigma} \nabla_{\rho}
({R^{\nu}}_{\sigma}-({1}/{4}) {\delta^{\nu}}_{\sigma} R )
/{\sqrt{-g}} , $
which is traceless and covariantly conserved \cite{Dese:82}. It
would be a non-trivial task to find the general black hole solutions
for the third-derivative-order equations. However, there is a
trivial vacuum solution, e.g., the BTZ solution because it has a
constant curvature $R=-6/l^2$ and hence satisfies the equation
(\ref{eom:EHGCS}) trivially, with the vanishing Cotton tensor $C^{\m
\n}= \epsilon^{\mu \rho \sigma} \nabla_{\rho}
{R^{\nu}}_{\sigma}/{\sqrt{-g}}=0$ \cite{Kalo:93}.

Even though, for the BTZ solution, there are no contributions in the
equations of motion (\ref{eom:EHGCS}) from the GCS term, there are
some important non-trivial effects as follows. First, their bare
mass and angular momentum are shifted as
\begin{eqnarray}
 M=m+\hat{\be} j/l, ~~ J=j+ \hat{\be}l m \label{MJ:GCS},
\end{eqnarray}
respectively \ci{Garc:03,
Krau:06,Solo:06}. When $|\hb | \ra \infty$, this becomes the exotic
black hole system in Sec. II B \ci{Solo:06,Park:0602,Park:0608}. For
a finite $\hb$, this interpolates between the ordinary and exotic
BTZ black holes depending on its value, as can be easily seen from
the relation
$ M^2 -J^2/l^2 =(1-\hat{\be}^2) (m^2-j^2/l^2)$:
For the small values of coupling $|\hb|<1$, the usual inequality is
preserved, i.e., $M^2  \geq J^2/l^2$; however, for the large values
of coupling $|\hb|>1$, one has an {\it anomalous} inequality with an
exchanged role of the mass and angular momentum as $J^2/l^2 \geq
M^2$; and also at the critical value $|\hb|=1$, the modified mass
and angular momentum are ``always'' saturated, i.e., $M^2 =
J^2/l^2$, regardless of inequality of the bare parameters $m$ and
$j$.

The second nontrivial effect is that, as a result of the shifts in
(\ref{MJ:GCS}), the black hole entropy has a term proportional to
${\cal A}_-$, as well as ${\cal A}_+$ \ci{Solo:06}. But as in the
case of exotic black holes in Sec. II B, there  are problems in the
conventional approaches of Refs. \ci{Krau:06,Solo:06,Saho:06}: The
usual approaches, which agree with the results from the Euclidean
method of conical singularity \ci{Solo:06} and Wald's \ci{Saho:06},
with the usual Hawking temperature $T_+$ and angular velocity
$\Om_+$ as in (\ref{T:BTZ}), give the entropy
\begin{eqnarray}
\label{S:old2:GCS}
 S_{GCS}=\frac{{\cal A}_+}{4 G \hbar}+\hb \frac{{\cal A}_-}{4
G \hbar}
\end{eqnarray}
with the usual first law
$
 \delta M=\Omega_+\delta J + T_+ \delta S_{GCS},
$
but the second law is not guaranteed in general, especially for the
large $\hb$. Recently, I have argued that (\ref{S:old2:GCS}) is
valid only for $|\hb|<1$, but the correct entropy for $|\hb|>1$,
satisfying the second law, be \ci{Park:0602,Park:0608}
\begin{eqnarray}
S_{GCS(\new)}=\hat{\ep} \left(\frac{{\cal A}_-}{4 G \hbar}+\hb
\frac{{\cal A}_+}{4 G \hbar} \right) \label{S:new:GCS}
\end{eqnarray}
with $\hat{\ep}=sign(\hb)$.

Here, I note that, in contrast to the case of the exotic black
holes, the appearance of the inner-horizon term is inevitable either
in (\ref{S:old2:GCS}) or (\ref{S:new:GCS}). This seems to imply that
the usual black-hole ``hologram'' picture \ci{Beke:73,tHoo:93} needs
to be corrected with the GCS term. In particular, the concept of
black-hole entropy as the measure of information about a black-hole
interior which is accessible to an exterior observer might not be
valid in our case; but, it is not quite clear whether this implies
the probing of the black-hole interior by the GCS action
\ci{Solo:06}.

Now, let me explicitly demonstrate that how the second law be
satisfied by (\ref{S:old2:GCS}) and (\ref{S:new:GCS}), depending on
the values of $\hb$. Actually, this can be easily demonstrated by
combining the previous results in Sec. IIA and Sec. IIB since our
system is a mixture of them, basically. First, for the small
coupling $|\hb|<1$, one has the usual inequality $M \geq J/l$ as in
(\ref{ieq:BTZ}). So the area increasing theorem of (\ref{area:BTZ})
applies, for the quasi-stationary process satisfying $\de M \geq \de
J/l \geq 0$ with $T_+>0, \Om_+>0$, but now for the combined areas,
\begin{\eq}
\label{area:+GCS}
\de ({\cal A}_+ +\hb {\cal A}_-)&=&\f{4 \hbar G}{T_+} ~(\de M-\Om_+ \de J) \no \\
 &\geq& \f{4 \hbar G}{T_+}  \left( 1-\f{r_-}{r_+}
 \right) \de M \geq 0.
\end{\eq}
This result implies that the increment $\de {\cal A}_+$ from
(\ref{area:BTZ}) dominates some possible decrement $\de {\cal A}_-$
from (\ref{area:BTZ:inner1}), if there is, always such as the
combination ${\cal A}_+ +\hb {\cal A}_-$ is always increasing. This
implies the second law, due to the positive temperature $T_+$.

On the other hand, for the large coupling $|\hb|>1$, one has the
inequalities
$
 0 \leq  M \leq J/l
$
(for $\hb>1$) or
$
 0 \geq  M \geq J/l
$
(for $\hb<-1$).
Then, the area-increasing theorem, for the quasi-stationary process
satisfying $0 \leq  \de M \leq  \de J/l$,
applies
for the combined areas,
\begin{\eq}
\label{area:-GCS}  \de ({\cal A}_- +\hb {\cal A}_+)&=& \f{4 \hbar
G}{T_-}
~(\de M-\Om_- \de J) \no \\
 &\geq&  \f{4 \hbar G}{T_-}~ \left( 1-\f{r_+}{r_-}
 \right) \de J \geq 0
\end{\eq}
for $\hb >1$ and $\de J \geq 0$. Then, it is clear that the entropy
formula (\ref{S:new:GCS}) does satisfy the second law in this case;
the additional sign factor $\hat{\ep}=-1$ for $\hb <-1$ is necessary
in order to make the entropy increases definitely.

Finally, I note that the above results (\ref{area:+GCS}) and
(\ref{area:-GCS}) may be considered as a {\it generalized} area
theorem.
\\

\begin{subsection}
{The
Kerr black hole in four dimensions}
\end{subsection}

In Sec. IIA, I have studied the dynamics of the inner and outer
horizons for the BTZ black hole in three dimensional anti-de Sitter
space. But, the behaviors seem to be
quite generic if there are, at least, two horizons and symmetries
between them. In this section, I study the $four$-dimensional Kerr
black hole for the vanishing cosmological constant and show that
this shows quite similar behaviors as the BTZ black hole in Sec.
IIA.

To this end, I first start from the Kerr black hole solution in
(3+1) dimensions, in the standard form \ci{Myer:86}
\begin{eqnarray}
\label{Kerr} d s^2 =-\frac{\rho^2 \ga}{\Sigma^2} dt^2
+\frac{\Sigma^2}{\rho^2} \mbox{sin}^2 \theta \left(d \phi-\frac{a
\mu r  }{ \Sigma^2} dt \right)^2 +\frac{\rho ^2}{\ga} d r^2 + \rho^2
d \theta^2 ,
\end{eqnarray}
where
\begin{eqnarray}
&&\rho^2 =r^2 +a^2 \mbox{cos}^2 \theta, ~~\ga=r^2 + a^2 -{\mu r} , \nonumber \\
&&\Sigma^2=\rho^2 (r^2 +a^2) +{\mu r} a^2 \mbox{sin}^2 \theta.
\end{eqnarray}
The ADM mass and angular momentum are given by
\begin{\eq}
\label{MJ:Kerr}
 M=\f{\m}{2 G},~~J=M a,
\end{\eq}
respectively. The inner and outer horizons are determined by $\ga
=0$ as
\begin{eqnarray}
r_{\pm}&=&\f{\mu \pm \sqrt{\mu^2 -4 a^2}}{2} \no \\
 &=&MG \pm \sqrt{M^2G^2 -a^2} .
\end{eqnarray}
Note that the mass, which needs to be positive, and angular momentum
satisfy the inequality
\begin{\eq}
\label{M>J:Kerr}
 GM^2 \geq J
\end{\eq}
or equivalently
\begin{\eq}
\label{m>a:Kerr}
 \m \geq 2 a
\end{\eq}
in order that the horizons exist, similarly to (\ref{ieq:BTZ}).
The symmetry between the inner and outer
horizons, in the metric, would be manifest by noting that
\begin{\eq}
\m =r_{+} +r_{-},~~ a^2=r_{+} r_{-}.
\end{\eq}

Then, it is not difficult to see that the quantities in
(\ref{MJ:Kerr}) satisfy the following identities, for an arbitrary
variation $\de$,
\begin{\eq}
\de M -\Om_+ \de J=\f{T_+}{4 G \hbar} \de {\cal A}_+, \\
\de M -\Om_- \de J=\f{T_-}{4 G \hbar} \de {\cal A}_-,
\end{\eq}
where
\begin{eqnarray}
&&T_{\pm}=\left. \frac{\hbar \kappa}{2 \pi}
\right|_{r_{\pm}}=\frac{\hbar (r_{\pm} -r_{\mp})}{4 \pi (r_+ +r_-)
r_{\pm}}, \no \\
&& \Omega_{\pm}
 =\frac{a}{r_{\pm}^2+a^2}, \no \\
&&{\cal A}_{\pm}=4 \pi (r_{\pm}^2 +a^2)
\end{eqnarray}
are the parameters for temperatures, angular velocities and the
horizon areas, respectively. But, as I have noted in Sec. IIA,
there is $no$ physical meaning of $T_-$ and $S_-={\cal
A}_-/4G \hbar$ as the characteristic temperature and the entropy,
due to the lack of the second law for $S_-$.

The demonstration of the area theorem or the second law is similar
to that of Sec. IIA by considering the quasi-stationary process
satisfying
\begin{\eq}
\label{dM>dJ:Kerr}
  \de M \geq \f{1}{\m} \de J \geq 0,
\end{\eq}
such as (\ref{M>J:Kerr}) or (\ref{m>a:Kerr}) is not violated by the
perturbation. Then I find, for $T_+>0, \Om_+>0$ and $\de M\geq0$,
that
\begin{\eq}
\label{area:+Kerr}
\de {\cal A}_+&=&\f{4 \hbar G}{T_+} ~(\de M-\Om_+ \de J) \no \\
 &\geq& \f{4 \hbar G}{T_+} ~( 1-\m \Om_+ )
  \de M ~\geq~ 0,
\end{\eq}
where I have used
$\Om_+
\leq {1}/{\mu} $,
from (\ref{m>a:Kerr}). This demonstrates the Hawking's area theorem
for the restricted class of perturbation (\ref{dM>dJ:Kerr}).

On the other hand, for the inner horizon, I find that
\begin{\eq}
 \label{dAm:Kerr}
\de {\cal A}_-&=&\f{4 \hbar G}{ T_-} ~(\de M-\Om_- \de J) \\
 &\leq& \f{4 \hbar G}{ T_-} ~ ( 1-\m \Om_- )
  \de M
\end{\eq}
for $T_-<0$, i.e., $r_+ >r_- (>0)$ and $\de M \geq0$. Note that the
upper bound of $\de {\cal A}_-$ is positive definite due to the
lower bound
$\Om_-
\geq {1}/{\mu}, $
in contrast to the upper bound of $\Om_+$;
but this is very similar to the situation in BTZ black hole, i.e.,
(\ref{Om-}). From (\ref{dAm:Kerr}), one can easily demonstrate the
decreasing for ${\cal A}_-$, by considering $\de M\geq0,\de J=0$
without violating (\ref{dM>dJ:Kerr})\footnote{If I use the null
energy condition for the infalling matters, i.e., $\de M- \Om_+ \de
J \geq 0$, which is satisfied by (\ref{dM>dJ:Kerr}) also, I obtain
$\de {\cal A}_+ \geq 0$ generically \ci{Jaco:95}, but
(\ref{dA-<0:Kerr}) changes as $\de {\cal A}_- \leq 16 \pi G \m \de
M$. But here $\de M$ needs not be positive. (See footnote 2 for the
comparison.)},
\begin{\eq}
\label{dA-<0:Kerr} \de {\cal A}_-=\f{4 \hbar G}{ T_-} ~
  \de M ~\leq~0.
\end{\eq}
This shows that $\de {\cal A}_-$ is ``not'' positive
(semi-)definite\footnote{ A similar behavior in the
Reissner-Nordstrom black hole was first pointed out by G. Kang
\ci{Kang:06}.} such as $S_-={\cal A}_-/4G \hbar$ can not be
identified as an entropy. Finally, I note that the inequality
$
 \de {\cal A}_- \leq  \de {\cal A}_+
$
is still valid, as in the BTZ case, though it needs some algebra.\\

\begin{section}
{Concluding remarks}
\end{section}

I have demonstrated, using the quasi-stationary process, the
Hawking's area increasing theorem for the outer horizon and some
$possible$ decreasing for the inner horizon in various black holes
which include the recently discovered exotic black holes and BTZ
black holes in higher derivative/curvature gravities as well as the
usual BTZ and the Kerr black hole in four dimensions.
I have also demonstrated a generalized area theorem by combining
those of the outer and inner horizons.

The proposed entropies agree exactly with the statistical entropy
from the CFT analysis which computes the entropy directly,
independently of the first law
\ci{Park:0602,Park:0608,Park:0609}. But a difficult problem of the
new entropy formulae is that it requires {\it unusual}
characteristic temperature parameters
which differ from the usual Hawking temperature $T_+$,
as well as the unusual angular velocity parameters $\Om_-$, from the
``assumed'' first law of thermodynamics. The negative-valued
temperature {\it might} be understood from the existence of the
upper bound of masses, analogous to the spin systems \cite{Kitt:67},
but the very meanings of
the unusual temperature and angular velocity parameters in the
Hawking radiation \ci{Hawk:78} are not quite clear. This raises the
question whether the first law is really satisfied for those
anomalous black hole systems\footnote{In the derivation of the mass
formula, used in this paper, one uses the derivatives of the Killing
vector which necessarily contain derivatives of metric. Thus, the
very validity of the mass formula near the inner horizon might be
questionable due to some divergences in the gradients of metric
perturbations.}; if the first law is not necessarily valid, it would
be hard to justify the calling an increasing-area-law a second law,
from the thermodynamics only. Another independent analysis, like as
the statistical entropy computation, would be crucial in that
situation \ci{Park:02,Kang:04}.


Finally, I note that, as far as the area theorem for the outer
horizons is concerned, the usual derivations via the Raychaudhuri's
equation give the same result as the quasi-stationary process of
this paper since the same $vacuum$ Einstein equation, regardless of
the details of the actions, is satisfied for all the black hole
solutions in this paper,
though this would not be true in general. But its generalization to
the inner horizon does not seem to be so straightforward. It would
be interesting to clarify this question.\\

\section*{Acknowledgments}

This work was inspired by the discussion with Gungwon Kang. I thank
him for the  discussion. I also thank Sean Hayward, Hideki Maeda,
Alex B Nielsen for useful correspondences. This work was supported
by the Science Research Center Program of the Korea Science and
Engineering Foundation through the Center for Quantum Spacetime
(CQUeST) of Sogang University with grant number R11 - 2005- 021.

%
%
\newcommand{\J}[4]{#1 {\bf #2} #3 (#4)}
\newcommand{\andJ}[3]{{\bf #1} (#2) #3}
\newcommand{\AP}{Ann. Phys. (N.Y.)}
\newcommand{\MPL}{Mod. Phys. Lett.}
\newcommand{\NP}{Nucl. Phys.}
\newcommand{\PL}{Phys. Lett. }
\newcommand{\PR}{Phys. Rev. }
\newcommand{\PRL}{Phys. Rev. Lett.}
\newcommand{\PTP}{Prog. Theor. Phys.}
\newcommand{\CQG}{Class. Quant, Grav.}
\newcommand{\hep}[1]{ hep-th/{#1}}
\newcommand{\hepg}[1]{ gr-qc/{#1}}
\newcommand{\bi}{ \bibitem}

\end{document}